\documentclass[preprint,showpacs,amsmath,amssymb]{revtex4}
\usepackage{graphicx}
\usepackage{color}

\begin{document}
\title{How dsDNA breathing enhances its flexibility and instability on short length scales}
\author{O-chul Lee}
\email{lee572@postech.ac.kr}
\affiliation{Department of Physics, Pohang University of Science
and Technology, Pohang 790-784, Republic of Korea}
\author{Jae-Hyung Jeon}
\email{jae-hyung.jeon@ph.tum.de}
\thanks{Present address: \emph{Physik Department (T30g), Technical University of Munich,
James-Franck Strasse, Garching 85373, Germany}}
\affiliation{Department of Physics, Pohang University of Science
and Technology, Pohang 790-784, Republic of Korea}
\author{Wokyung Sung}
\email[corresponding author:]{wsung@postech.ac.kr}
\affiliation{Department of Physics, Pohang University of Science
and Technology, Pohang 790-784, Republic of Korea}

\date{\today}

\begin{abstract}
We study the unexpected high flexibility of short dsDNA which
recently has been reported by a number of experiments. Via the
Langevin dynamics simulation of our Breathing DNA model, first we
observe the formation of bubbles within the duplex and also forks
at the ends, with the size distributions independent of the
contour length. We find that these local denaturations at a
physiological temperature, despite their rare and transient
presence, can lower the persistence length drastically for a short
DNA segment in agreement with experiment.
\end{abstract}

\pacs{87.14.gk, 87.15.A-, 87.15.hg}

\maketitle

The DNA is a double helix of two single-stranded (ss) backbone
chains paired by complimentary bases via hydrogen-bonding and
further stabilized by stacking interaction between adjacent
base-pair (bp) planes \cite{1.Nelson}. Owing to recent advances of
single molecule techniques, intensive studies have been done on
structural changes and mechanical behaviors of double-stranded
(ds) DNA constrained by external forces and twists. The continuum
elastic model, called wormlike chain (WLC) model has been very
useful in analytically describing the micron-scale conformations
and elastic response of such DNA \cite{2.Mako}. The persistence
length, the measure of segmental orientation correlation, is about
50 nm (equivalently, about 150 bases along contour) for dsDNA. On
the contrary numerous biological facts suggests DNA loops more
readily on much shorter length scales. Indeed, Cloutier and Widom
showed that the DNA has much higher cyclization probability than
predicted by a WLC of persistence length of 50 nm
\cite{3.Cloutier}. Also, Wiggins $et~al.$ showed the DNA on short
length scale has an elastic behavior distinct from that of WLC
\cite{4.Wiggins}, while Yuan $et~al$. very recently have reported
the persistence lengths are as short as 11 nm for DNA fragments
consisting of $10\sim20$ base-pair (bp) \cite{5.Yuan}.

In this paper we demonstrate that the higher flexibility of dsDNA
emerges on shorter scales indeed due to local denaturation.
Because of the large initiation energy, the fraction of the open
bases is less than 1\% at the physiological temperature, and, once
formed, bubbles decay shortly in the order of 50 $\mu$s
\cite{7.5}, seemingly little affecting the DNA stability. We show
that despite their rare and transient presence the bubbles give a
drastic enhancement of the flexibility as the chain gets shorter.
For a very short duplex fragment, another type of local
denaturation, namely, forks at the free ends are entropically
favorable, dominating over the bubbles to enhance the flexibility,
and eventually driving the duplex to unbind into two single
strands.

It was suggested and estimated that the baseflips \cite{5.Yuan}
and kinks \cite{YanMarko} can enhance the bending flexibility and
looping probability. Based on a simple two state model, the
looping probability was evaluated using transfer matrix method \cite{YanMarko}.
Although all these calculations are suggestive of their
significance, the bubbles have not yet been explicitly accounted
for with regard to their size distributions and realistic
energetics.

\begin{figure}
\includegraphics[width=0.5\textwidth]{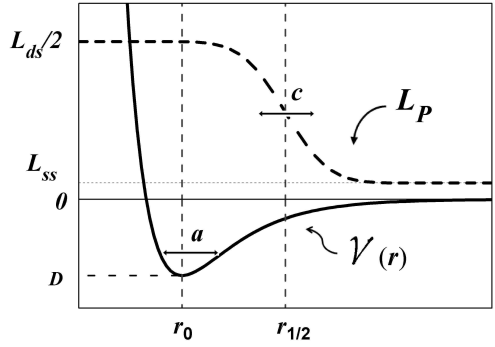}
\caption{A schematic figure showing the profile of persistence length $L_p$ per single strand (dashed curve), along with the Morse potential $\mathcal{V}(r)$ (bold curve) as function of bp distance $r$.}
\label{figure:lp}
\end{figure}

We consider a homogeneous dsDNA as the duplex of two interacting
single strands described by the effective energy
\begin{equation}
\mathcal{H}=\mathcal{H}_1+\mathcal{H}_2+\mathcal{V}_{12}.
\end{equation}
The $\mathcal{H}_i$ is the elastic energy of the single strand $i$
(=1 or 2), which, in a discrete representation, is given by
\begin{equation}
    \mathcal{H}_i=\sum_{n=2}^{N-1}\frac{\kappa}{2}(\textbf{r}_{n-1}^{(i)}-2\textbf{r}_{n}^{(i)}+\textbf{r}_{n+1}^{(i)})^{2}
  +\sum_{n=1}^{N-1}\frac{k}{2}(|\textbf{r}_{n+1}^{(i)}-\textbf{r}_{n}^{(i)}|-b)^{2},
\label{1}
\end{equation}
where $\textbf{r}_n^{(i)}$ is the three-dimensional position
vector of $n$th bead in strand $i$ ($n=1,2,\cdot\cdot\cdot,N$) and
$b$ ($\cong0.34$ nm) is an average distance between neighboring
beads within the ss. The first term accounts for bending energy
with the bending modulus $\kappa b^3$ for the ss, which is the
related to its persistence length $(L_{p})$ via $\kappa
b^3=L_{p}k_BT$ \cite{9.J-Y KIM}. The second term, the stretching
energy of each strand, is introduced to impose chain
inextensibility condition. An appropriate value of $k$ is
numerically found by matching simulation of either force-extension
curves or the mean end-to-end distances of ssDNA resulting from
this model, with corresponding theoretical expression of the
inextensible WLC without the stretching term. The
$\mathcal{V}_{12}$ is the pairing energy between complimentary
bases. To describe the bp openings due to thermal excitation,
namely, thermal bubbles and forks, the interaction is represented
by $\mathcal{V}_{12}=\sum^{N}_{n=1}\mathcal{V}_n(r_n)$ where
\begin{equation}
\mathcal{V}_n(r_n)=De^{-(r_n-r_{0})/a}[e^{-(r_n-r_{0})/a}-2]
\end{equation}
is the Morse potential \cite{9.5 M. Peyrard, Theo},
$r_n=|\textbf{r}^{(1)}_n-\textbf{r}^{(2)}_n|$ is the distance
between $n$th bp, $r_0$ and $a$ are the bond distance and range
which correspond to mean and fluctuation of DNA diameter
respectively, and $D$ is the potential depth. Whenever bps are
unbound, the duplex becomes no more than two single strands with
net persistence length as short as twice of $L_{ss}= 1\sim4$
nm~\cite{13.J. B. Mills}, while for bound bps it takes $L_{ds}=50$
nm, the persistence length of long DNA, due to stacking
interaction in the double strand. In order to include the stacking
and destacking that cause such variation of the persistence length
in the model, we consider $\kappa(n)=L_p(n)k_BT/2b^3$, and propose
the persistence length per single strand $L_p(n)$ takes the form
\begin{equation}
L_{p}(n)=L_{ds}/2-(L_{ds}/2-L_{ss})\tilde{\theta}(r_{n-1}-r_{1/2})
\tilde{\theta}(r_{n}-r_{1/2})\tilde{\theta}(r_{n+1}-r_{1/2}).\label{lpvar}
\end{equation}
Here $\tilde{\theta}(r)=[1-\hbox{erf}(r/c)]/2$ is a steplike form
function which smoothly increases from $0$ to $1$ over the width
$c$ and $r_{1/2}$ is the distance at which $L_p$ is $(L_{ds}/2+L_{ss})/2$ (see Fig.~\ref{figure:lp}).

$L_p$ depends on three consecutive bp distances naturally as it
represents the cooperativity of the stacking interaction~\cite{lpcomment}.
The variation of $L_p$ depending on the bp distance $r$ is
schematically illustrated in Fig.~\ref{figure:lp}: When all of the
three consecutive bps are outside the range $r_{1/2}$, namely they
are unbound one other, $L_p$ takes the single-stranded persistence
length $L_{ss}$ while as any one of the three are bound, $L_p$ is
reduced to $L_{ds}/2$. The $c$ is comparable to the Morse
potential width $a$ since the $L_p$ varies due to the bp unbinding.

Using this energy model, we simulate the dynamics and equilibrium
distribution of the bp distance, via the Langevin equation,
\begin{equation}\label{2}
    \Gamma\frac{d}{dt}\textbf{r}^{(i)}_{n}(t)=-\frac{\partial\mathcal{H}}{\partial\textbf{r}^{(i)}_{n}}+\boldsymbol{\xi}^{(i)}_{n}(t),
\end{equation}
where frictional coefficient per base is $\Gamma=6\pi \eta
R=1.88\times10^{-11}$ kg s$^{-1}$ with $\eta=$0.001 kgm$^{-1}
s^{-1}$ for viscosity of water and $R=1$ nm for effective radius
of nucleotide. $\mathbf{\xi}_{n}$ is the Gaussian and white noise
satisfying $\langle\xi_{n \alpha}^{(i)}\rangle = 0, \langle\xi_{n
\alpha}^{(i)}(t) \xi_{n'\alpha'}^{(j)}(t')\rangle= 2 \Gamma
k_BT\delta_{ij}\delta_{n n'} \delta_{\alpha\alpha'}\delta(t-t')$
for each Cartesian component $\alpha$ and $\alpha'$. The values of
potential depth and range, $D=0.07$~eV, $a=0.05$~nm are chosen so
that the 300-bp DNA fragments have the melting temperature 350~K
\cite{9.J-Y KIM}.

\begin{figure}
\includegraphics[width=0.8\textwidth]{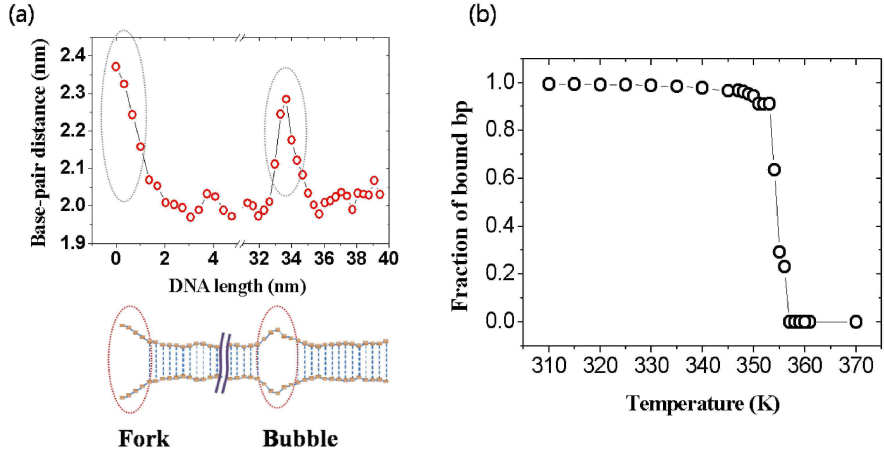}
\caption{(a) A snapshot of bp distance in a duplex of
300-bp ($\sim 100$ nm) long obtained from simulations of the Breathing DNA model.
The shape of duplex thickness therefrom drawn manifests breathing, formation of bubble
in the middle of the duplex and forks at the ends.
(b) The simulation data on fraction of bound bp in the 300-bp long duplex vs. temperature.
It shows a sharp transition into  denaturation state at melting
temperature ($\approx$ 350 K). At physiological temperature ($\sim$310
K), the fraction less than 1\% is denatured (unbounded) either
as bubbles or forks.
}\label{figure:example}
\end{figure}

In our simulation, we employed 50 nm for the persistence length of
dsDNA ($L_{ds}$) and 4 nm for that of ssDNA ($L_{ss}$) with the
stretching constant $k=41$ pN/nm. For the parameters in $L_p(n)$,
$r_{1/2}$ was chosen to $(r_0+r_c)/2=2.1$~nm where $r_c$ is the
cutoff distance at which a bp is regarded as unbound in the
simulation, and $c$ is 0.045 nm~\cite{9.J-Y KIM}. We checked that
small variation of the parameters values does not affect
significantly the main result of our simulation.

Figure \ref{figure:example} shows a snapshot profile of
equilibrated bp distance along the contour of 300 bp long or about
100 nm, with free ends, and the fluctuating DNA thickness of about
2 nm therefrom constructed. The figure shows indeed the local
denaturation exist in the forms of bubbles within the contour and
fork at the ends. From the simulation we find the fraction of
bound pairs undergoes a sharp transition from near unity to zero
in agreement with experiment \cite{9.3} at the melting
temperature, which is about 350 K. It indicates that at the
physiological temperature (310 K), the fraction of the bp that
forms the local denaturation is less than 1\%.

\begin{figure}
\includegraphics[width=0.8\textwidth]{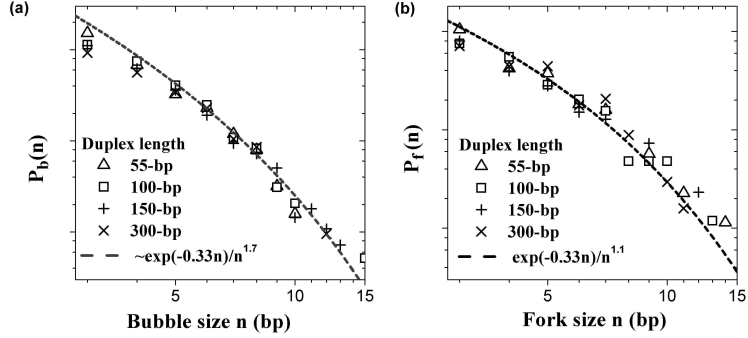}
\caption{(a) The bubble size distribution $P_b(n)$ at 310 K for contour length
 $L$ = 55, 100, 150, and 300 bp. (b) The fork size distribution $P_f(n)$ at 310 K
 for the same lengths.}\label{figure:dis}
\end{figure}

Shown in the Fig. \ref{figure:dis} are the size distributions of
the bubbles $(P_b(n))$ and forks $(P_f(n))$ for the various
contour lengths of the DNA fragments. The distribution of the
bubble size with $n$ bp open is remarkably independent of its
position and the DNA length. For relatively large bubbles it
follows the Poland-Sherega form \cite{11.J-H Jeon}
\begin{equation}
P_b(n)\sim \frac{e^{-n\Delta}}{n^{\alpha}},
\label{3}
\end{equation}
where $\alpha$ is the statistical factor for a loop formation and
$\Delta$ is the average energy in $k_BT$ to unbind a bp. These
factors are independent of the length and given by those for a
long length, $\alpha=1.7$, $\Delta=0.33$, at 310 K \cite{9.J-Y
KIM}. The average size of the bubbles for various contour lengths
is 1.7 bp at 310 K. This is in agreement with the
Peyrard-Bishop-Dauxois model simulation \cite{9.6}. The
length-independence of their distribution and small average size
implies that the thermal bubbles are transiently excited due to
short-range elastic interaction in unconstrained DNA. This is to
be contrasted with the dsDNA under mechanical constraints, where
the average bubble size is found to be much larger depending on
the (negative) twist as well as the contour length \cite{9.7}. The
distribution of $n$ bp open at a free end is given by
\begin{equation}
P_f(n)\sim \frac{e^{-n\Delta}}{n^{\beta}},
\label{4}
\end{equation}
where $\beta=1.1$, $\Delta=0.33$ are also independent of the
contour length within the errors. The average size of the forks is
1.9 bp. Since $L_{ss}$ is 4 nm in these simulations, they are
semiflexible forks with $\beta$ much larger than that of the
flexible forks \cite{9.75}

\begin{figure}
\includegraphics[width=0.5\textwidth]{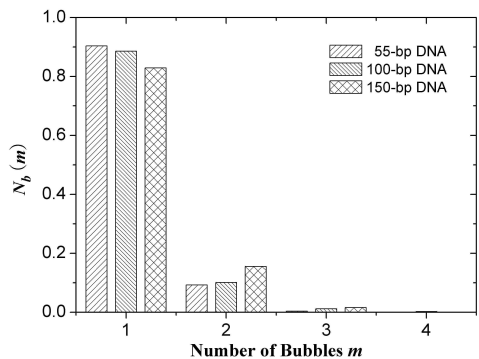}
\caption{The relative probability of finding $m$ bubbles within three DNA lengths at 310~K.}
\label{figure:bcf}
\end{figure}

Last, we obtain the $N_b(m)$, the relative probability of finding
$m$ bubbles \emph{simultaneously} in a DNA of given length. The
result obtained from simulation is presented in Fig.~\ref{figure:bcf},
showing that the single bubble occurrence becomes predominant as
the DNA length gets shorter. This is mainly due to the large
energy cost of bubble initiation; namely, once a bubble is formed,
increasing its size is energetically more favorable than opening a
new bubble elsewhere. Within the short DNAs of our interest, the
number of bubbles, if they exist at all, can be regarded to be
unity in a good approximation.

\begin{figure}
\includegraphics[width=0.5\textwidth]{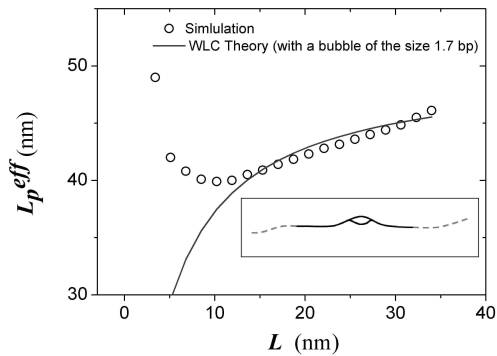}
\caption{Effective persistence length of the DNA fragments of the length $L$
taken from inside a 300 bp long DNA (empty circles) This is compared
with theoretical curve (line) obtained assuming a single bubble
of the size given by the mean, 1.7 bp.}\label{figure:data1}
\end{figure}

To focus on the effects of bubble (excluding the end fork effect)
on the overall duplex persistence length, we consider an ensemble
of fragments with given contour length $L$ randomly taken
\emph{from inside} a 300-bp long ($\sim100$ nm) dsDNA at a
physiological temperature. The square of duplex end-to-end
distance, $\textbf{R}_{ee}^2$, is defined as arithmetical average
of the end-to-end distances of the two single strands. The thermal
average of its square $\langle\textbf{R}_{ee}^2\rangle$ is taken
over the ensemble of the duplex with a given $L$. Suppose that the
duplex to be a WLC with a uniform persistence length $L_p^{eff}$,
we use the well-known relation~\cite{9.8}
\begin{equation}
\langle\textbf{R}^2_{ee}\rangle=2L^{eff}_pL-2(L^{eff}_p)^2[1-e^{-L/L^{eff}_p}].
\end{equation}
Comparing this with simulation data, we evaluate $L_p^{eff}$ of
our duplex for various values of the contour length $L$. The
result shown in Fig.~\ref{figure:data1} indicates that $L_p^{eff}$
decreases from the value 50 nm to 40 nm as the $L$ decreases to 10
nm. It reflects the enhanced flexibility of the duplex, which we
attribute to the presence of thermal bubbles. Along the DNA length
shorter than 50 nm, if any, a single bubble is most likely to
exist, as shown in $N_b(m)$, Fig.~\ref{figure:bcf}. If the single
bubble survives as the contour length decreases, its mean size
does not change as discussed earlier, consequently yielding larger
flexibility.

To support this argument quantitatively, we note that the mean
end-to-end distance is given by
$\langle\textbf{R}^2_{ee}\rangle=\int_0^L \int_0^L ds ds'
\langle\textbf{t}(s)\cdot\textbf{t}(s')\rangle$~\cite{9.8}, where
$\textbf{t}(s)$ is the unit tangent vector at the arclength $s$
from an end of the duplex. The tangent correlation function is
given by $\langle\textbf{t}(s)\cdot\textbf{t}(s')\rangle =
\exp\{-|s'-s|/\emph{L}_{ds}\}$ if they are the positions within ds
region and $\langle\textbf{t}(s)\cdot\textbf{t}(s')\rangle =
\exp\{-|s'-s|/2\emph{L}_{ss}\}$ if $s$ and $s'$ are the points
within the bubble region. Integrating the correlation function along the
contour with a bubble of the mean size, 1.7 bp, we evaluate the
$\langle\textbf{R}_{ee}^2\rangle$. Since a bubble occurs with
equal probability along the contour we further average the mean
square end-to-end distance over all possible position of the
bubble, and relate it with the effective persistence length, which
is shown by the curve in Fig.~\ref{figure:data1}.

For the length larger than 10 nm, this analytical theory agrees
with simulation result remarkably, suggesting that the persistence
length reduction is indeed due to a single bubble. As the length
decreases below  that, this curve departs much from
the simulation result for the length shorter than that. It is
because even a single bubble is unlikely to occur within such a
short length so that the persistence length sharply rises to the
double strand value. For a long DNA fragment, on the other hand,
only a bubble is found in most cases with its average size fixed
independently of DNA length, which results in no significant
perturbations on the persistence length.

\begin{figure}
\includegraphics[width=0.7\textwidth]{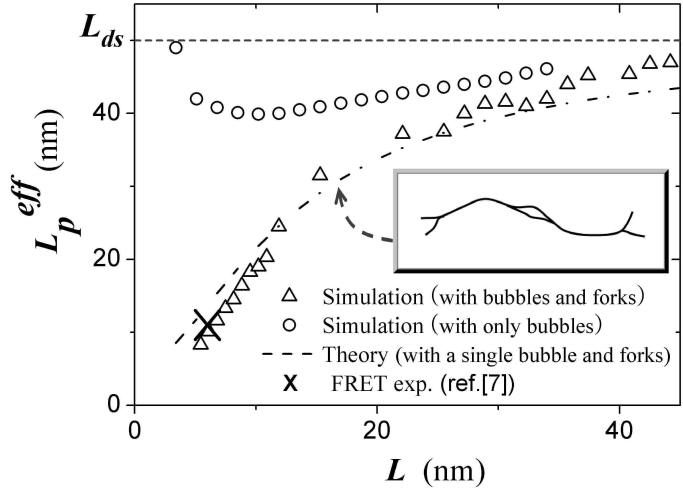}
\caption{Effective persistence length of the
fragment of DNA with contour lengths $L$,
where bubbles as well as forks exist (triangle).
It is in a good agreement with the WLC theoretical curve with a bubble of the size given as its average, 1.7 bp and two forks of the size given as its average, 1.9 bp, and with the  results of ref. \cite{5.Yuan} obtained by  a fluorescence resonance energy transfer (FRET) experiment.}
\label{figure:data2}
\end{figure}

\emph{For the short duplex with free ends}, we also have
investigated the mean end-to-end distances of DNA fragments length
shorter than 50 nm (147 bp) at 310~K. The persistence length
calculated in a similar way from these data is shown by triangles
in Fig.~\ref{figure:data2}. It clearly shows that $L_p^{eff}$, in
this range of the contour length, are shorter than $L_{ds}=50$~nm
but converges to the value as the contour length increases. The
persistence length in this case is shorter than that of the duplex
of the same contour length without the free ends discussed before
(indicated by circles). This can be ascribed to the additional
form of local denaturation, i.e., the forks at the free ends. To
support this quantitatively, we analytically calculated the
effective persistence length of the duplex with a single bubble of
the size 1.7 bp and two forks of the size 1.9 bp at the ends by
integrating the correlation function and following the procedure
as before. The close agreement of the theoretical curve with the
simulation evidences additional influence of the forks; the two
forks combines with and increasingly dominates over a bubble, to
enhance flexibility, as the duplex gets shorter.

In ref. \cite{5.Yuan}, the effective persistence length of dsDNAs
of contour length $15\sim21$ bp (equivalently $5\sim7$ nm) was
determined to be $11\pm2$ nm by a fluorescence resonance energy
transfer (FRET) experiment, which is marked by a cross in Fig.
\ref{figure:data2}. They considered the buffer conditions where
the persistence lengths of non-interacting single stranded DNA
have the range of the values 2.7-3 nm~\cite{13.J. B. Mills}. In a
duplex, however, the two unpaired single strands are subject to
steric and electrostatic repulsions, which may enhance this ss
persistence length effectively to about 4 nm. Indeed, our values
of $L_p^{eff}$ estimated from simulation with $L_{ss}$(=4 nm) are
in a good agreement with this experimental value.

Simulating the Breathing DNA model that incorporates locally
fluctuating persistence lengths depending upon the bp distances
along the contour, we find that the distributions of bubbles and
forks are nearly independent of the contour length, reducing the
effective persistence length of the duplex for the short lengths.
The forks dominate the bubbles in enhancing duplex flexibility
and inducing unbinding transition.

\acknowledgments

This work was supported by NCRC and BK 21. We thank L. A. Archer
for a valuable communication.


\begin{thebibliography}{99}
\bibitem{lpcomment} A similar description was also
introduced in the Peyrard-Bishop-Dauxois model \cite{Theo} where
the stacking energy is described by a harmonic energy with a
bp-distance dependent spring constant, $k(r_n,r_{n-1})=K_0(1+\rho
\exp[-\alpha(r_n+r_{n-1}-2r_0)])$.
\end{thebibliography}
\end{document}